# Molecular lighting goes less powering

*Ranjeev Kumar Parashar[a], Suchita Kandpal[b], Prasanta Bandyopadhyay[a], Mainak Sadhukhan[a], Rajesh Kumar[b],\* and Prakash Chandra Mondal[a]\**

[a]*Department of Chemistry, Indian Institute of Technology Kanpur, Uttar Pradesh-208016, India*
[b]*Department of Physics, Indian Institute of Technology Indore, Simrol-453552, India*

\*Email: rajeshkumar@iiti.ac.in (R.K.); pcmondal@iitk.ac.in (P.C. M.)

**Abstract**

The present era has seen tremendous demands for low-cost electrochromic materials for visible-region multicolor display technology, paper-based, flexible, and wearable electronic devices, smart windows, and optoelectronic applications. Towards this goal, we report large-scale polyelectrochromic devices fabricated on rigid to flexible ITO substrates comprising novel anthracene containing viologen, (1,1'-bis(anthracen-9-ylmethyl)-[4,4'-bipyridine]-1,1'-diium bromide, abbreviated as $AnV^{2+}$), and polythiophene (P3HT). Interestingly, the devices show three states of reversible visible color in response to the applied bias, sub-second to second switching time (0.7 s/1.6 s), and high coloration efficiency (484 cm$^2$/C), longer cycling stability up to 3000 s (10$^3$ switching cycles). Thanks to the anthracenes moieties introduced to viologen that inhibit formation of undesired dimer of cation radicals formed in response to the applied bias, otherwise it would hamper the devices reconfiguration. The devices are fully characterized, and electrochromic performances are ensured by bias-dependent UV-Vis, and Raman spectroscopy. The fabricated electrochromic devices are tested with the commercially available low-cost cells to perform, which is highly desired for practical applications. The computational study facilitates the understanding of experimental results. The alternating current (AC)-based electrical impedance spectroscopy reveals that P3HT facilitates reducing charge transfer resistance of the devices. Our work shows CMOS compatibility and one of the best-performing devices that could pave the way for developing cost-effective flexible, and wearable electrochromic devices.

**Keywords:** viologen, P3HT, reversible redox change, coloration efficiency, flexible electrochromic device

# Introduction

The search for novel organic electrochromic materials continues due to their easy synthesis, solution processability, low-cost, ultra-high vacuum-free device fabrication, thermal stability, and low-bias operational in multi-state color switching in the entire visible range (400 -780 nm).[1–3] Electrochromism (EC) refers to the reversible, distinct color changes occurring in the materials in response to bias applied to the devices. This phenomenon is undoubtedly considered as s a breakthrough in the realm of low-cost, flexible organic-compounds-based electrochromic materials for fabricating solid-state devices. Further, electrochromic devices (ECDs) become more widespread when they portray polychromic properties, high coloration efficiency (CE), and sub-second switching time that plays vital roles in energy consumption.[4,5] Thanks to flexibility of ECD device, by virtue of it, it can be exploited in the wearable electronic gadgets, display, but its development is in the infant stage. So far, various electrochromic materials including inorganic metal oxides ($WO_3$), perovskites, nanomaterials, coordination compounds, and organic conducting polymers, and metallo-supramolecular polymers are well-explored as active color changing components in the electrochromic devices.[6–12] Despite a plethora of accomplishments in inorganic-based ECDs, their widespread use is impeded by factors such as the high-cost of inorganic metal oxides, low-coloration efficiency, high energy consumption, long-term stability, and requiring sophisticated techniques including cleanroom facilities or ultra-high vacuum depositions employed to produce defect-free thin films which are both tedious and expensive. In contrast, the organic electrochromic materials such as conducting polymers such as polyanilines, polypyrroles, polythiophenes and viologens (4,4′-bipyridine derivatives) have emerged as an appealing candidate due to mesmerizing properties such as ease of synthesis, vivid color, efficiently structure tuning, scalability, low-production cost.[13–17] Organic materials show promising solution-processable; therefore, their homogeneous and desired thickness of thin films can be prepared by employing relatively straightforward techniques like drop casting, spin coating, and electrografting allowing to form homostructure to heterostructures highly desirable for suitable electrochromic device fabrication.[18,19] Viologens, among the electrochromic organic materials have considered the utmost attention due to their ease of chemical structures control and modification, tunable optoelectronic properties, wider ranges of thermal stability, bistable redox states, room temperature solubility, air-stability, large-area surface coating and easy characterization, can withstand larger electrochemical potential window, coloration both in solution and solid-state, and long-range charge transport ability.[13,20,21] Depending on the substituents, viologens may exist in three distinct redox forms: dication (pale yellow/red-colored or colorless), mono-radical-cation (violet-blue/green), and diradical (generally colorless).[13]. In an application to a bias, the dicationic form ($V^{2+}$) undergoes two different forms, a cation radical ($V^{+\bullet}$) and a neutral species ($V^0$), and their radical cations exhibit exceptional stability. Due to their individual electronic properties, the various redox states exhibit naked-eye observable optical properties (absorption/transmittance). However, it is challenging to suppress the dimeric formation between the cation radicals, driven by radicals reactivity, and π-π interaction between the rings. Such dimeric form is irreversible, therefore it's not possible for the devices to show reconfigurability. To address this issue, through our smart design, we prepared viologen that contains two anthracene moieties that are oriented *trans* to each other, suppressing dimeric formation but does not hamper the co-planarity, thus enabling delocalization of free radicals within bipyridine rings formed during one electron reduction process. The present work demonstrates rigid-to-flexible electrochromic device fabrication using the novel anthracene viologen (acts as electron acceptor) mixed with P3HT (acts as hole acceptor) assembled by means of the classical flip-chip method. The devices show fast and reversible multi-color switching, high-coloration efficiency, longer cyclic stability that even performs well when the devices are forced to bend at the angle of 40º made on the flexible substrate. We further attempt to run the fabricated devices by a low-cost commercial Duracell (1.5 US$ for 10 cells of 1 V) and performs color switching phenomena without the need for an expensive Potentiostat, a step toward practical applications.

**Result and discussion**

A solid-state single crystal structure of anthracene viologen (AnV$^{2+}$, CCDC†: 2214301, **Table S1**.) is shown in **Fig. 1a** used for electrochromic device fabrication on rigid and plastic transparent conductive oxide substrates. The compound was prepared in a three-step reaction (See SI for synthesis and characterization data, **Fig. S1–S4**). The AnV$^{2+}$ was crystallized in the monoclinic space group P2$_1$/c with a crystal density of 1.585 g cm$^{-3}$ at 100(2) K and C18-C18a, C18–C19 bond lengths are found to be 1.484(6) and 1.392(4) Å, respectively. Two anthracene rings in the viologen are oriented in *trans* fashion, while a torsion angle of 180º indicates the excellent planarity of two pyridiniums. The molecular electrostatic potential (MESP) plot for AnV$^{2+}$ is mapped over the electron density surface (0.001 a.u. iso-value, **Fig.1b**). Since this is a dicationic system, the high positive values of MESP demonstrating bipyridinium moieties are electron deficient. Note that the whole 4,4'-bipyridinum moiety has electron density spread homogeneously over two pyridine rings which are at par with the extended conjugation described above. The viologen used for the ECD device fabrication shows high thermal stability, which reveals only 5% weight loss at 255ºC (**Fig. S5**), thus can sustain high-temperature.

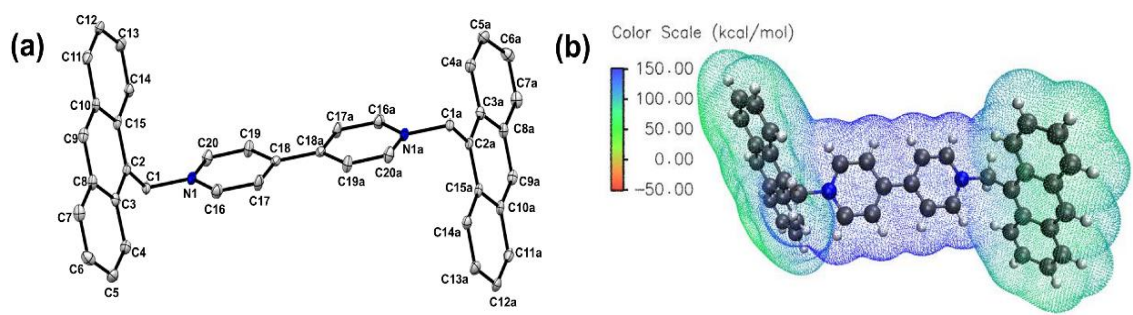

**Fig. 1.** (a) Single-crystal structure of AnV$^{2+}$ with thermal ellipsoids is drawn on 50% probability (H atoms, and counter anions are excluded for clarity), (b) electrostatic potential mapped on the electron density surface of AnV$^{2+}$ (iso-value = 0.001 a.u.) of AnV$^{2+}$.

The UV-vis absorption spectrum of AnV$^{2+}$ recorded at room temperature (300K) in DMSO showing four characteristic absorptions centered at 338, 358, 375, and 395 nm (**Fig. 2a**). The emission spectrum ($\lambda_{ex}$ = 375 nm) exhibits the maximum emission in the visible regime (~400–500 nm, **Fig. 2a**), with the $\lambda_{em, max}$ =

413 nm. UV–vis spectrum of a thin film of AnV$^{2+}$ deposited on ITO shows similar absorption but with a redshifted of the λ$_{max}$ and exploited the data for calculation of optical bandgap which appeared at 2.6 eV deduced from the Tauc plot (**Fig. 2b-c**). Cyclic voltammogram of AnV$^{2+}$-coated ITO-films exhibit two reversible single-electron redox processes occurring at the modified ITO-electrode/electrolyte interface where the first reduction occurs at -0.68 V, and the second one at -1.12 V Vs. Ag/AgNO$_3$ reference electrode. The first cathodic peak ensures formation of cation radical, AnV$^{+•}$, and the second one corresponds to neutral form (AnV) which can be reconfigured upon reversing the bias (**Fig. 2d**). The reversible behavior is diffusionless mass transport process, an indicative of fast electrochemical kinetics, very similar to the ferrocene redox chemistry.[22] The current density (mA/cm$^2$), even without double layer contribution (capacitive current) for both the reduced species, shows relatively higher than that of oxidized species (I$_{pc}$/I$_{pa}$ >1), ensuring a strong tendency to uptake the electron from the Fermi level (E$_F$) of ITO to its unoccupied molecular orbitals. The voltammogram shows two distinct reduction peaks well-separated by 450 mV and two oxidations are separated by 430 mV, indicating efficient generation of of bi-stable redox species without overlapping each other. Such redox features are highly desirable for achieving controllable multi-color states, while the lower onset cathodic waves (-0.6 V, and -1 V) indicate low-energy redox phenomena suitable for energy-efficient electrochromic devices integration using AnV$^{2+}$ as an active electrochromic material. Fast, and reversible electron transfer kinetics that drive the redox and chemical reversibility at the electrode/electrolyte interfaces, along with four well-defined redox signals even at narrow potential ranges, are quite encouraging for band gap determination, (d) cyclic voltammogram of AnV$^{2+}$ thin films made on ITO (used as working electrode) using 0.1 M TBAP electrolyte, scan rate = 50 mV s$^{-1}$.

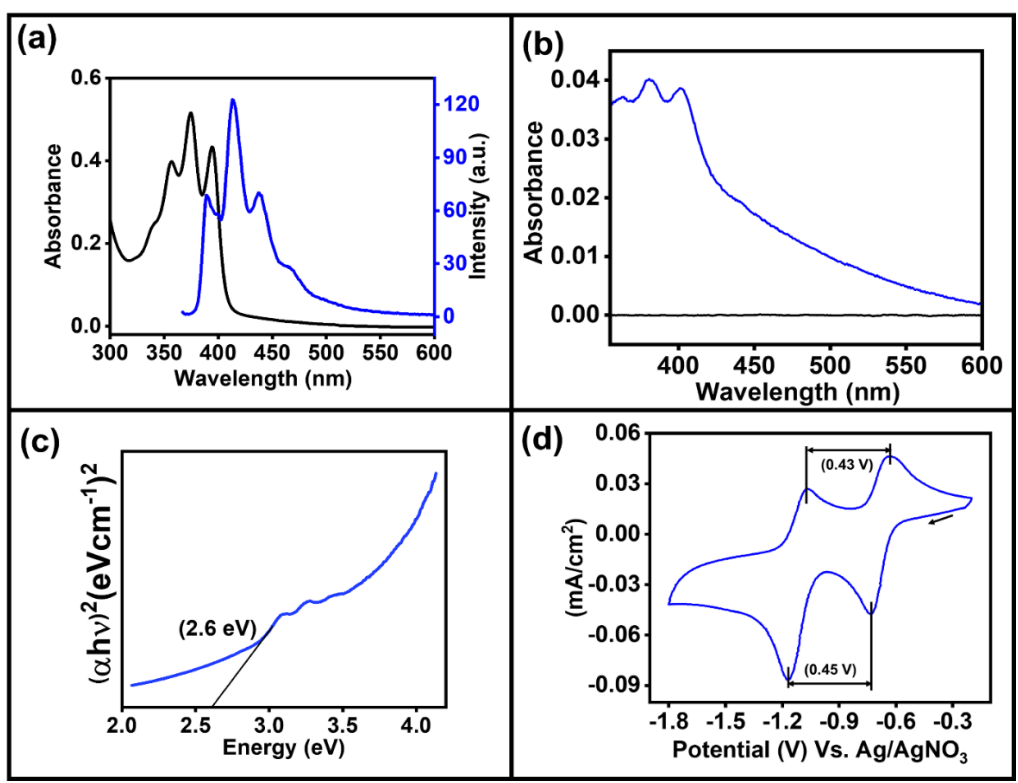

**Fig. 2**. (a) UV-vis absorption (Black) and emission spectrum (Blue) of the anthracene viologen (AnV$^{2+}$), (b) solid-state UV–vis spectrum recorded on thin ITO films (blackline indicates ITO baseline), (c) Tauc Plot used for optical

bias-driven optical phenomena, which motivate us to design and fabricate solid-state ECDs. UV-vis absorption and CV data are employed to calculate the HOMO-LUMO energy gap at 2.6 eV, which is closer to the DFT calculated value (see computational support section).

Field-emission scanning electron microscopy (FE-SEM) image of AnV$^{2+}$ reveals micro-structures of AnV$^{2+}$ on ITO showing fiber-like morphology with varying diameters which play crucial roles in meeting long-range charge transport, and we take the advantage of it in high-performance electrochromic functions (**Fig. 3a**). The P3HT films rather show uniform and dense with streamer-shaped features (**Fig. 3b**). The cross-sectional FE-SEM images of AnV$^{2+}$, and P3HT films are employed to measure an average thickness of 12.8 ± 0.3, 2.1±0.2 µm, respectively (**Fig. S6**). X-ray photoelectron spectra (XPS) of C 1s core level show three deconvoluted peaks with binding energies (BE) appearing at 286.2, 284.8, and 284.6 eV assigned to C=N, –CH$_2$–, and aromatic rings (C–C/C–H), respectively (**Fig. 3c,** see survey spectrum in **Fig. S7**). N 1s spectra could be fitted into two BEs of 402.0, 399.6 eV attributed to dicationic, and neutral N atoms in the viologen (**Fig. 3d**). We couldn't observe formation of monocationic radicals during the exposure of high-energetic X-ray photoelectron to the thin films of AnV$^{2+}$, and cross-checked with EPR spectroscopy. However, some groups observed photo-electron driven charge transfer occurred in the viologen derivatives during the XPS measurements.[23,24] Our system unequivocally indicates structural robustness, and long term stability. Further, to check the stability of our novel viologen, a solution AnV$^{2+}$ in DMSO was exposed to sunlight and UV irradiation independently which does not alter the photophysical properties, thus formation of dimerization, or photo-mediated oxidation can be ruled out (SI section **S8**). The core-level spectra of Br 3d exhibit two BEs at 68.4, and 67.4 eV, ascribed to $3d_{3/2}$, $3d_{5/2}$, respectively which implies that Br⁻ exist as counter ions in the $AnV^{2+}$, ensure no formation of C-Br bonds (**Fig. 3e**). XPS valence band spectrum is used to determine the valence band maxima deduced at 2.1 eV (**Fig. 3f**).

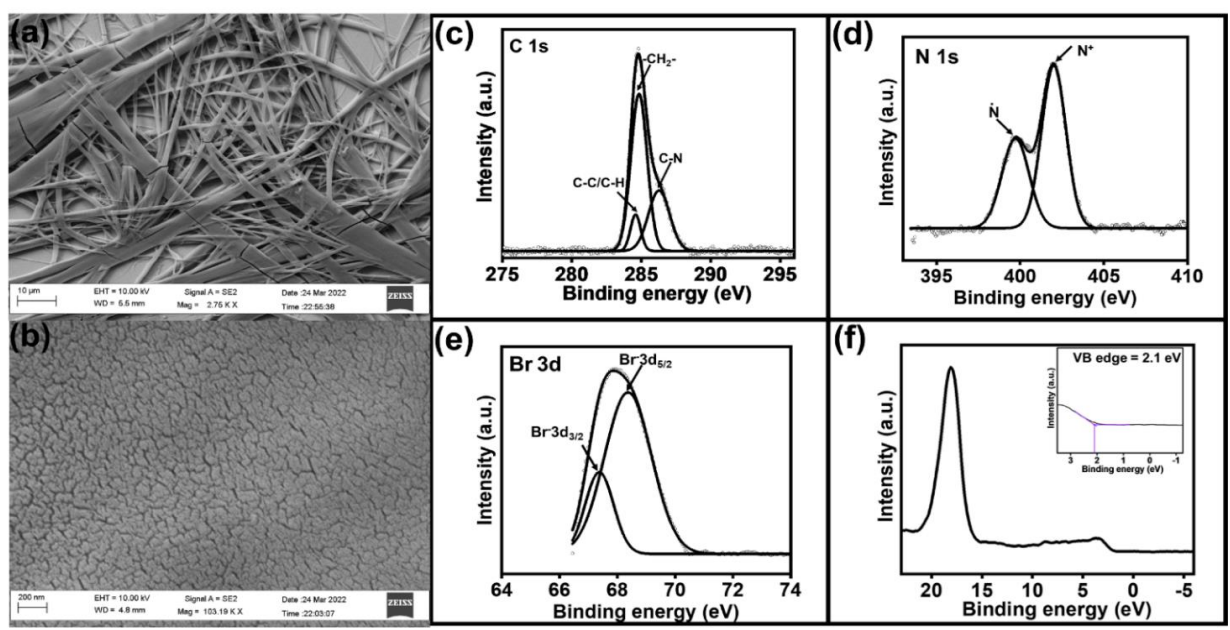

**Fig. 3.** FE-SEM images of (a) AnV$^{2+}$ thin films, (b) P3HT thin films prepared on ITO. XP spectra of (c) C 1s, (d) N 1s, (e) Br 3d core level, and (f) valence band edge from XP spectra.

## Solid-state electrochromic devices: Fabrication and testing

Solid-state electrochemical devices (ECDs) were fabricated using $AnV^{2+}$ and P3HT, where these films were sandwiched between two freshly cleaned ITO electrodes using a classical flip-chip method. One important aspect is to mention here that in addition to DMSO, our synthesized anthracene viologen is soluble in methanol, and in solvent mixture like methanol: water (1:1, V/V), methanol: acetonitrile (2:1, V/V) and acetonitrile: water (2:1, V/V), thus provides an excellent opportunity for easy thin films preparation. Preparation of solution, thin films followed by electrochromic devices fabrication on ITO and PET substrates have been discussed in the supporting information (see Section **S9**). The rigid ECDs fabricated on ITO (both bottom and top electrodes, actual photograph is **Fig. S8**) exhibit multiple colors switching between magenta and blue with their intermediate (transparent) state at a relatively lower bias than the blue state (**Fig. 4a-c**). The as-prepared devices appear magenta, which is the combined color of $AnV^{2+}$ and P3HT mixed layers, when no electric field was applied. When a bias of +1 V is applied to the P3HT-deposited ITO electrode, the device turns transparent due to the oxidation of the neutral P3HT, forming polaronic states. On further increasing the bias from +1 V to +1.8 V, the device changes its color to blue. At this applied bias, $AnV^{2+}$ layers (connected to the –Ve terminal) get reduced to convert to cation radicals, which is blue in color. The anthracene moieties help to exist as monocationic radicals, a strong indication of structural stability and such stable formation is important to switch into other states and return to the initial states for performing many cycles without any deterioration. In-situ bias-dependent absorbance spectra of the devices were recorded to see the spectral changes. The unbiased device (black curve, **Fig. 4d**) absorbs the maximum near the green band ($\lambda_{max}$ = 525 nm) of the visible spectrum, meaning transmission of the remaining two primary colors (red and blue), producing the magenta color of the device. At +1V, the device almost equally absorbs the entire visible wavelength, thus appearing transparent (red curve, **Fig. 4d**). On further increasing the voltage from +1 V to +1.8 V, the overall absorption band shifts towards lower energy and starts absorbing in the green-red region (blue curve, **Fig. 4d**), leaving only 'blue' color to the device. As bias polarity is reversed from positive to negative direction, at -1.8 V, the device color switches back to its original magenta color (green curve, **Fig. 4d**) with a recovery of 95±2%. Since the device shows multiple color switching (magenta – transparent – blue) in different but low biases, thus various other essential parameters, to quantify device performance, like response time, cycle life, current response, and coloration efficiency in two different bias conditions (1V, 1.8V) are evaluated. For switching from magenta to transparent, a rectangular pulse of voltage ±1 V of 10 s duration (5s each polarity) was applied to operate the device, and the cyclic response in absorbance (corresponding to 515 nm) was recorded. Single switching cycle shows that the device takes 0.8s/2.7s to switch its color between magenta and transparent (**Fig. 5a**) and such fast chromism is desired for practical applications, thanks to our "all-organic" module help to achieve such excellent performances at low-operational bias. Significantly, less variation in absorbance in the repetition of such switching steps up to 3000s or 1000 switching cycle shows a great stability/cyclability of the devices (**Fig. 5b**). Further, the efficiency has been estimated after 3000 cycles run of the electrochromic devices for a time of 10000s which comes out to be 419 cm$^2$/C (coloration efficiency plot, **Fig. S9**), comparable with the value of the freshly prepared devices (~464 cm$^2$/C). The comparable efficiencies confirm that the device does not deteriorate much with continuous operation. The electrochromic coloration efficiency (CE) which is a crucial electrochromic parameter for evaluating ECD performance was using the equation (1)

$$\eta = \Delta OD/Q \qquad (1)$$

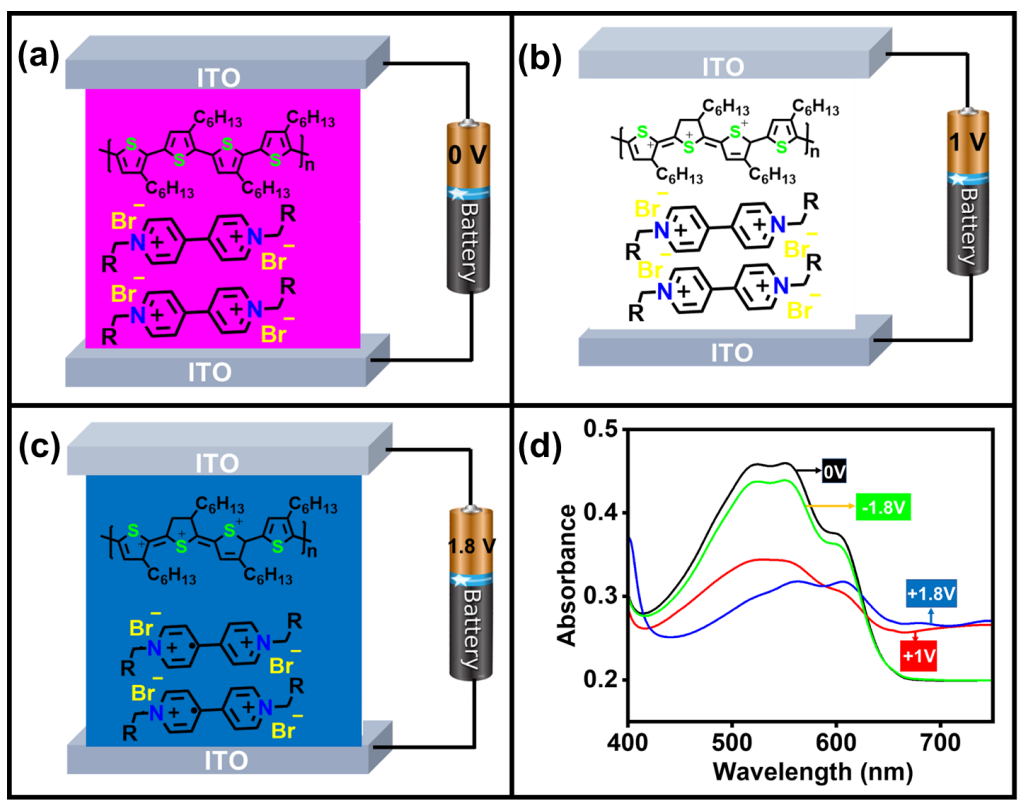

**Fig. 4.** (a-c) Schematic illustration of the electrochromic device showing different visible colors at different applied biases and (d) in-situ bias-dependence UV-vis spectra of the devices.

Where $\eta$ is the CE in cm$^2$ C$^{-1}$, $\Delta OD$ is the variation in optical density, and Q (C/cm$^2$) is the charge density. Herein, the coloration efficiency of the device was calculated from the slope of the curve plotted between the change in optical density ($\Delta OD$, obtained from **Fig. 5a** and the charge density (Q) obtained from **Fig. 5c** and $\eta$ calculated at 409±6 cm$^2$/C (**Fig. 5d**). The switching response of the device was also tested towards the response to higher bias ranges at a rectangular voltage pulse ± 1.8V of 10s duration (5s each bias polarity). It produces fast responses, as the devices take only 0.7s/1.6s to reach the absorption up to 90% of the maximum value (**Fig. 5e**). The device also shows enhanced stability with a minimal change in absorbance over the timescale of 3000s (equivalent to 1000 switching) corresponding to the chronoamperometric response (**Fig. 5f**). Using the current response and absorbance modulation plot (**Fig. 5g**), a coloration efficiency of 464±11 cm$^2$/C has been estimated for magenta-blue switching from the device which is one of the best performances molecular electrochromic devices (**Fig. 5h**). This also establishes that both the color switching regimes (magenta/transparent & magenta/blue) are efficient and fast with excellent stability. The plot of current density Vs. time (**Fig. S10**) employed to estimate charge density and found to be $5.3\times10^{-4}$ $C/cm^2$ for the reduction process and $3.0\times10^{-4}$ $C/cm^2$ for the oxidation process. Additionally, the cyclic voltammograms of individual electrodes deposited with $AnV^{2+}$, P3HT, and device fabricated thereafter have also been performed (**Fig. S11**). In situ-bias dependent Raman spectra of the devices were also recorded in ON and OFF states to confirm the coloration switching mechanism (**Fig. S12**). The black curve corresponding to the OFF state shows two distinct peaks at 1379 $cm^{-1}$ and 1444 $cm^{-1}$, signifying the presence of neutral P3HT. In an applied bias (+1.8V, ON state), these two distinct peaks come closer at 1376 $cm^{-1}$ and 1414 $cm^{-1}$ (blue curve), and a peak appeared at 950 $cm^{-1}$ (marked with *) confirmed the formation of polaron and cation radicals which match well with the previous reports.[25–27] With reversing the bias (-1.8 V), Raman signals come to their initial states, showing the reinstallation of neutral P3HT in the device, similar to the bias-dependent UV-vis studies with the devices. More discussion has been made in the supporting section (**SI, S14**). Interestingly, instead of using an expensive source measure unit or potentiostat, ECD was operated using low-cost Duracell (1.5 US$ for 10 cells) at the applied bias 1V and 1.8 V, showing similar electrochromic performances.

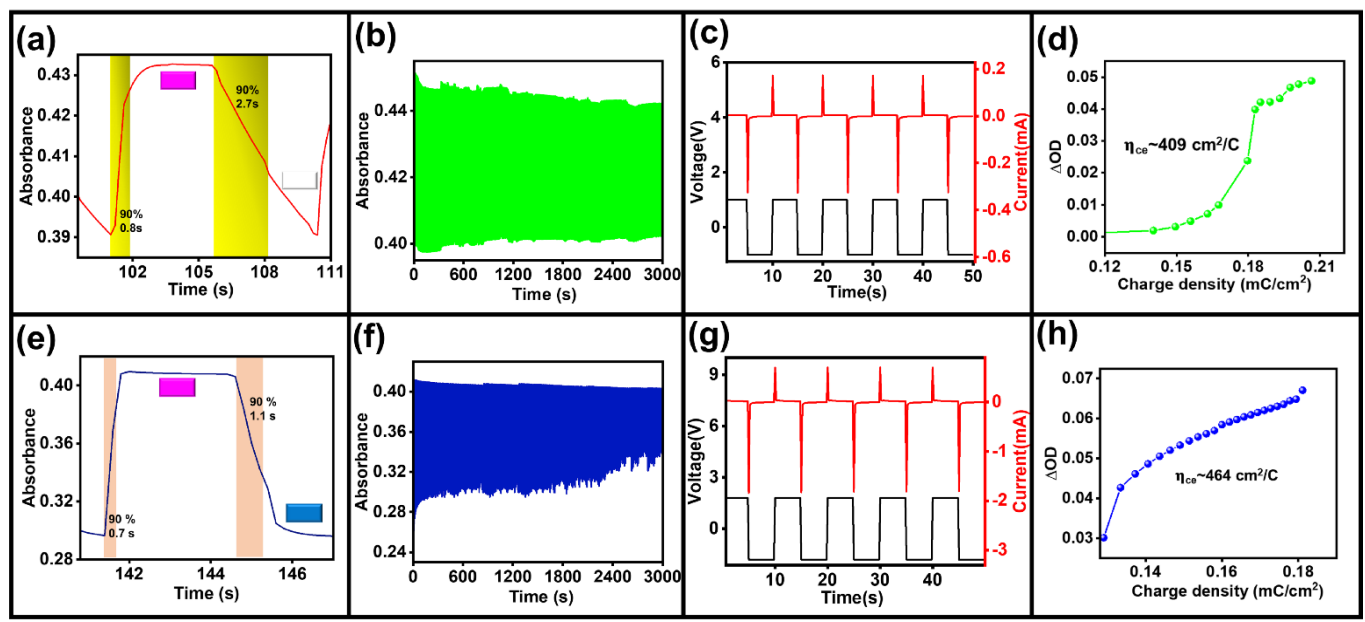

**Fig. 5.** (a) Single switching cycle of device between colored and bleached states in bias window (+1V to -1V), (b) multiple switching cycles, and (c) change in current through the device with time at fixed bias ±1V pulses and (d) estimation of coloration efficiency under ±1V switching. (e) response time of the device under ± 1.8V range, (f) multiple switching cycles up to 3000s, (g) current response of the device under ± 1.8V pulse, and (h) optical density variation as a function of charge density and measured the coloration efficiency.

To meet the demands for flexible electrochromic devices, we further fabricated the same on flexible substrate FTO coated on PET. The devices display a similar reversible three-state color switching from magenta to transparent to blue like solid-state devices, which is enthralling and a unique addition to the realm of electrochromic devices. Besides, flexible and large-area ECDs were also fabricated with dimensions 3 cm x 3 cm, and 5 cm x 5 cm area (**Fig. S13**). The flexible ECDs show excellent stability up to 3000s (1000 switching cycles) with minimal degradation in the absorbance value. Fast switching time, sub-second to second time-scale (0.9/2.8 s) was obtained, along with a high coloration efficiency of 434±6 cm$^2$/C (**Fig. S14**). Similarly, a rectangular pulse of ±1.8 V of 10 s duration (5 s for each polarity) was applied to the flexible ECDs that produces excellent electrochromic performance with excellent stability up to 3000s (1000 switching), switching time (1s/0.8s), and very high coloration efficiency 484±9 cm$^2$/C (**Fig. S15**). The flexible devices are bent at an angle of 40º which does not affect showing the high electrochemical performances (**Fig. S16**). To support that our ECDs are one of the best-performing devices, we compared the performance parameters of various viologen-based electrochromic devices reported earlier and compared them with ours (**Table S2**).

**Electrical impedance spectroscopy on electrochromic devices**

For practical applications, one must need to understand frequency-dependent electrical performances, as DC-based measurements are unable to produce crucial parameters such as uncompensated resistance ($R_{contact}$), charge-transfer resistance ($R_{ct}$), and double-layer capacitance ($C_{dl}$) or constant phase elements (CPE).[28] To achieve the goal, we performed electrical impedance spectroscopy (EIS) on the ECDs via two probes measurements (30 mV AC amplitude, frequency = 10 to $10^5$ Hz). Due to the inhomogeneity of the interface, we consider CPE rather $C_{dl}$, which is denoted as $Z_{CPE} = 1/Q\,(i\omega)^\alpha$ where α is the CPE exponent (lies between 0 and 1), and Q is the CPE parameter. We fabricated two different devices, ITO/AnV$^{2+}$/ITO (reference device), and ITO/AnV$^{2+}$/P3HT/ITO (ECD) to understand the roles of P3HT and its electrical property. The Nyquist plot (plot of imaginary vs. real components of impedance) for reference devices shows a steep rise at a lower frequency, indicating dominant capacitance behavior, while the ECDs show a bending with lowering the frequency (**Fig. 6a,c**). Charge-transfer resistance due to AnV$^{2+}$, and AnV$^{2+}$-P3HT are experimentally obtained as 6 MΩ, and 205 Ω, indicating that in presence of P3HT, the device behaves more conducting. Similarly, the uncompensated resistance is higher for reference devices than that of ECDs (35.5 KΩ and 66 Ω). However, it is evident that the Nyquist plot does not provide direct frequency response, but the Bode plot enables an easier visualization of the frequency variation to absolute impedance (|Z|). The Bode plot at the lower frequency of 10 Hz shows high resistance for ITO/AnV$^{2+}$/ITO, while the resistance of nearly five orders of magnitude less ($10^9$ Ω vs. $10^4$ Ω) was observed for ECD (**Fig. 6b,d**). At the lower frequency (longer time scale), capacitive reactance plays a significant role in the overall resistance of the devices. The above results are fitted to obtain equivalent circuit elements, where reference devices show almost identical interfaces, while ECDs reveal two different interfaces, ITO/AnV$^{2+}$, and ITO/P3HT (**Fig. 6e-f**). A loop formation with the ECD demonstrates that chemical reaction, due to the electron transfer between AnV$^{2+}$ and P3HT performing fast polyelectrochromic events.

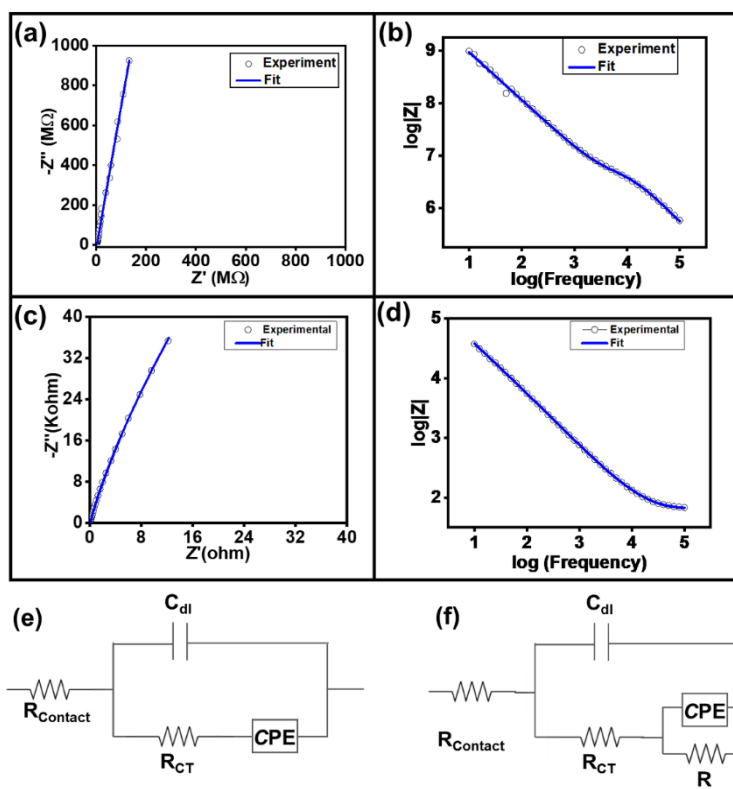

**Fig. 6.** (a, c) Nyquist plot for reference, and ECDs, (b,d) Bode plot for reference, and ECDs at the frequency range from 10 to $10^5$ Hz. Equivalent circuit model for (e) reference, and (f) ECDs.

**Computational support**

To understand the mechanism of charge transfer between $AnV^{2+}$, P3HT, and the ITO, we performed density functional theoretical computation. The initial structure of the $AnV^{2+}$ is taken from the crystal structure mentioned earlier. Counter-anion $Br^-$ does not have a significant effect on the structure and properties under consideration.[29] The $AnV^{2+}$ is successively reduced to radical cation $AnV^{+\bullet}$ and finally to the neutral form AnV. The structures of these complexes are optimized in DMSO solvent (**Fig. S17**). The computed HOMO-LUMO gap agrees very well with experimental results (**Fig. S18-19**). The pentameric structure for P3HT is used for the present computation (The justification is presented in **Fig. S20**, Table **S3**). The Fermi level of the ITO lies within the conduction band (-5.08 eV).[30] The energy landscape of frontier orbitals for all species involved in the charge-transfer process is presented in **Fig. 7**. In forward bias, the LUMO of the $AnV^{2+}$ accepts the incoming electron. The resulting $AnV^{+\bullet}$, having lower β-spin SUMO, accommodates another electron to produce AnV. The reverse bias triggers the opposite chain of events. The oxidation of AnV occurs by removing an electron from the HOMO, which further gets oxidized by ejecting an electron from α-spin SOMO, producing $AnV^{2+}$. All geometries are optimized by PBE0/def2-TZVP theoretical method in a vacuum and in DMSO (CPCM solvent model).[31–35] All visualizations and data processing are done with Multiwfn, VMD, and JMOL.[36–38] The $-C_6H_{13}$ groups of P3HT are replaced with $-CH_3$ for computational viability.

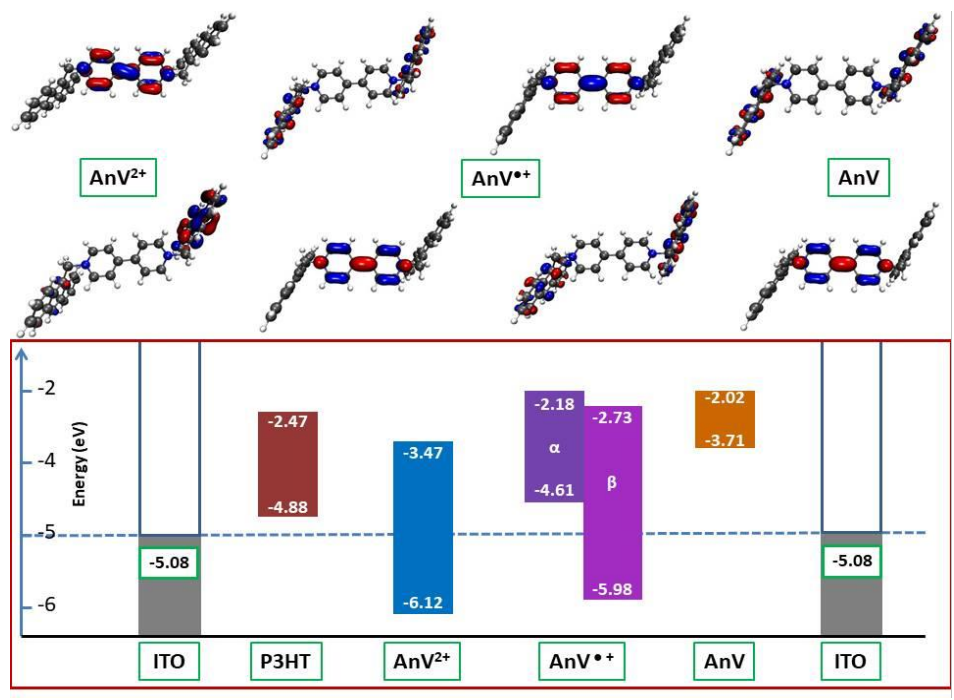

**Fig. 7**. Frontier molecular orbital energy landscape for the molecules involved in the electronic charge transfer pathways. The dashed horizontal line indicates the Fermi level of uncharged ITO electrodes.

## Conclusion

In a nutshell, flexible to rigid electrochromic devices are fabricated using a novel anthracene viologen as the cathodic electrochromic material, and P3HT as the anodic material showcasing outstanding devices properties in terms of coloration efficiency, sub-second switching time, low power consumption, multi-color switching, and excellent cycling stability. Our smart design strategy lies in the introduction of two anthracene moieties linked to the viologen ($AnV^{2+}$) which helps to retain stable cation radicals. Such chemical stability is crucial for the devices for reconfiguration when the bias polarity is reversed without destruction of coloration efficiency. We attempt here not to use metal-oxides-based electrochromic components that are known to problematic in many aspects, thus we provide here an "all-organic" approach to fabricate large-scale electrochromic devices. Considering the limitation of traditional DC-based measurements, AC-based measurements ensure internal charge transfer reducing the resistance of the devices and forming several redox-state to modulate the coloration in the visible region. Further, the inherent mechanism of the oxidation-reduction reaction in the device is also explained in terms of density functional calculations. The fabricated large-area, and flexible devices can be run by low-cost cells, without needing an expensive power supply. We anticipate that our innovative, and exciting work as "proof-of-concept" facilitates in fabricating prototypes of "all-organic" devices for power-efficient electronic displays, and wearable electronic gadgets in the near future.

## Acknowledgments

R.K.P., and S.K. thank CSIR, New Delhi, and IIT Indore for a Ph.D. fellowship. M.S. acknowledges a start-up research grant from IIT Kanpur (IITK/CHM/2018419). PB acknowledges an IPDF from IIT Kanpur.

MS acknowledges SERB grant SRG/2019/000369 for partial support. M.S. also acknowledges Debashree Manna for many helpful discussions. P.C.M. acknowledges financial support from Science and Engineering Research Board (Grant No. CRG/2022/005325), and Council of Scientific & Industrial Research, project NO.:01(3049)/21/EMR-II, New Delhi, India. The authors acknowledge IIT Kanpur for infrastructures and equipment facilities.